# A recent history of science cases for optical interferometry

Denis Defrère (ULiège, BE), Conny Aerts (ULeu, BE), M. Kishimoto (UKyo, JP), and P. Léna (LESIA, FR);

Contact email: ddefrere@uliege.be

Affiliations:
ULiège: Space sciences, Technologies and Astrophysics Research (STAR) Institute, Université de Liège, 19c Allée du Six Août, B-4000 Liège, Belgium
ULeu: Institute of Astronomy, KU Leuven, Belgium;
UKyo: Kyoto Sangyo University, Kyoto 603-8555, Japan;
LESIA: Observatoire de Paris-Meudon, Paris, France

Abstract: Optical long-baseline interferometry is a unique and powerful technique for astronomical research. Since the 1980's (with I2T, GI2T, Mark I to III, SUSI, ...), optical interferometers have produced an increasing number of scientific papers covering various fields of astrophysics. As current interferometric facilities are reaching their maturity, we take the opportunity in this paper to summarize the conclusions of a few key meetings, workshops, and conferences dedicated to interferometry. We present the most persistent recommendations related to science cases and discuss some key technological developments required to address them. In the era of extremely large telescopes, optical long-baseline interferometers will remain crucial to probe the smallest spatial scales and make breakthrough discoveries.

## 1. Introduction

The seed of many great discoveries in observational astronomy often stems from a scientific meeting, workshop or conference during which the blurry concept of an innovative instrument is discussed for the first time. For large optical interferometers, there are probably no better examples of such a success story than the VLTI (Léna 2005) and the CHARA array (ten Brummelaar et al. 2005). Although optical interferometry was first proposed in 1868 (Fizeau 1868) and led to its first scientific results around the turn of the twentieth century (see Michelson 1891 and Michelson & Pease 1921), it remained almost idle until 1975 when the idea of coherently coupling large telescopes was first proposed by A. Labeyrie during a Conference in Geneva on Optical telescopes of the future (Pacini et al. 1977). Today, as several interferometric facilities are reaching their maturity, we take the opportunity to review and summarize the best recommendations and conclusions established during past conferences about the future of optical interferometry. We discuss in particular the most persisting science cases that must be tackled by next-generation interferometric instruments and/or facilities.

Over the past decades, the astronomical community met at many occasions to discuss about the best science cases for optical interferometry and the adequate technological roadmap (see Figure 1 and the list of most workshops and conferences since 1994 here: http://iau-c54.wikispaces.com/Meetings). In Europe, the first meeting clearly addressing the science cases in the post-VLTI era occurred in Liège in 2004 (Surdej et al. 2004) and was followed one year later by a conference on the technology roadmap (Surdej et al. 2005). At the same time, the future of the VLTI was intensely discussed and various concepts were reviewed during a conference in Garching in 2005 on the power of optical interferometry (Richichi et al. 2005). In 2010, during the JENAM conference, the working group on the Future of Interferometry in Europe (FIE WG) defined the near-term priorities for the VLTI, a long-term vision in the ELT era, and guidelines for future facilities. The first step of the plan is now coming to fruition with the second-generation interferometric instruments bringing the VLTI capabilities and science to the next level (e.g. see GRAVITY first-light paper, Abuter et al., 2017).

In parallel, on the US side, nearly independent workshops came to similar conclusions regarding the best science cases for optical interferometry and the need for higher angular resolution. In particular, during two workshops held in Tucson (2006) and Socorro (2011) on the future of astronomy with long-baseline optical Interferometry, very high angular resolution was identified as the top priority and a key element toward a revolution in both planet formation and stellar physics. The study of energetic and interacting systems, including Active Galactic Nuclei (AGN), relativistic stellar systems, and binary systems with mass transfer was also listed as a second high priority. These science cases also emerged later from the Astro2010 decadal survey as top priorities. While international collaboration was often recognized during these meetings as essential for any future major interferometric facility, there was no consensus in the community on the best concept to choose and the lack of prospective vision was deplored at a few occasions. In addition, most discussions were generally focused on improving current facilities rather than on the need for a new international facility. This situation changed in 2013 during the OHP Colloquium "Improving the performances of current optical interferometers & future designs". During a round-table discussion between members of the EII, ASHRA, FRINGE and IF working groups (see acronym list in footnote[1]), it was concluded that direct imaging of the planet formation process at AU-scale radii can serve as a versatile science case of broad interest in the astronomical community, which at the same time is sufficiently focused to help developing the technical roadmap towards the next interferometric facility (Surdej and Pott, 2013). Shortly after the meeting this conclusion was distilled into the Planet Formation Imager (PFI) project (see other contribution in this volume).

To be complete and parallel to the now classical Fizeau-Michelson interferometry, it is worth mentioning here three potential aspects of evolution in optical interferometry in the future. First, there may be a possible revival of intensity interferometry, pioneered by Hanbury-Brown & Twiss in the 60s, then abandoned for its lack of sensitivity despite other great advantages, notably with respect to atmospheric phase degradations. The currently planned Cerenkov Telescope Array (CTA) will scatter over 1 square km a number of "light collectors", probably not adaptable to the purpose of optical interferometry but, in the future, large similar apertures could provide kilometric-base resolution on sufficiently bright objects, especially stars (Rivet et al. 2018, this same volume). Second, up-conversion of mid-infrared radiation to visible one may be considered, on the basis of recent work by Darré et al (2016), who coupled telescopes of the CHARA interferometer with up-conversion to the visible achieved at the focus of each telescope. Several advantages may arise from this : reduction of thermal emission and associated noise by taking the IR signal closer to the first optical surface, easy phase transport through classical single mode fibers, cheap detection of the fringe signal. Third, the new concept of densified pupil and hypertelescope, proposed by Labeyrie (1996), is abandoning the classical request made to an interferometer, namely to provide homothetic input-output pupils. By using extended diluted pupils, the gains in resolution, image quality and sensitivity are high and tests are progressing on the ground, while a space version (LISE) becomes conceivable (Labeyrie 1996, Labeyrie 2013).

---

[1] EII: European Interferometry Initiative, FIE: Future of Interferometry in Europe, ASHRA: Actions de Haute Résolution Angulaire (France), FRINGE: Frontiers of Interferometry center in Germany, IF: Interferometry Forum.

Also, it is worth noticing that the above considerations and most conferences, in the last two decades, dealt essentially with ground-based optical interferometry. Indeed, science cases would also be numerous and extremely interesting for an interferometer observing from space, as it was argued for as early as 1993 with the *Darwin* project submitted to the European Space Agency (Cockell et al. 2009) or, on the US side, with the Space Interferometry Mission (SIM, Marr 2006) and the Terrestrial Planet Finder Interferometer (TPF-I, Beichman et al. 2006). An optical interferometer in space, providing long integration times and high sensitivity, probably represents the must in the domain. But it would require expensive technologies, especially for a formation-flying concept, which becomes necessary for decametric or longer baselines. Such technique is today being successfully explored with the PRISMA mission of the Swedish Space Corporation and ESA's future Proba 3 mission. The issue "ground vs. space" was already discussed in the late 1980s, questioning an approval of the VLTI with its atmospheric limitations. Fortunately the VLTI was decided, and 30 years later no interferometric space mission, even an exploratory one, has yet emerged! Although it would be wise to make sure such an exploration would not be disregarded in the future, it seems sound today to focus on science cases with ground-based instruments, which could deal with them in the next two or three decades.

This paper is articulated around the conclusions and recommendations of the few key meetings and conferences mentioned above. The following section summarizes the most persistent recommendations related to science cases and mentions some key technological developments required to address them. More details on possible technological avenues for long-baseline interferometry are given in other papers of this volume.

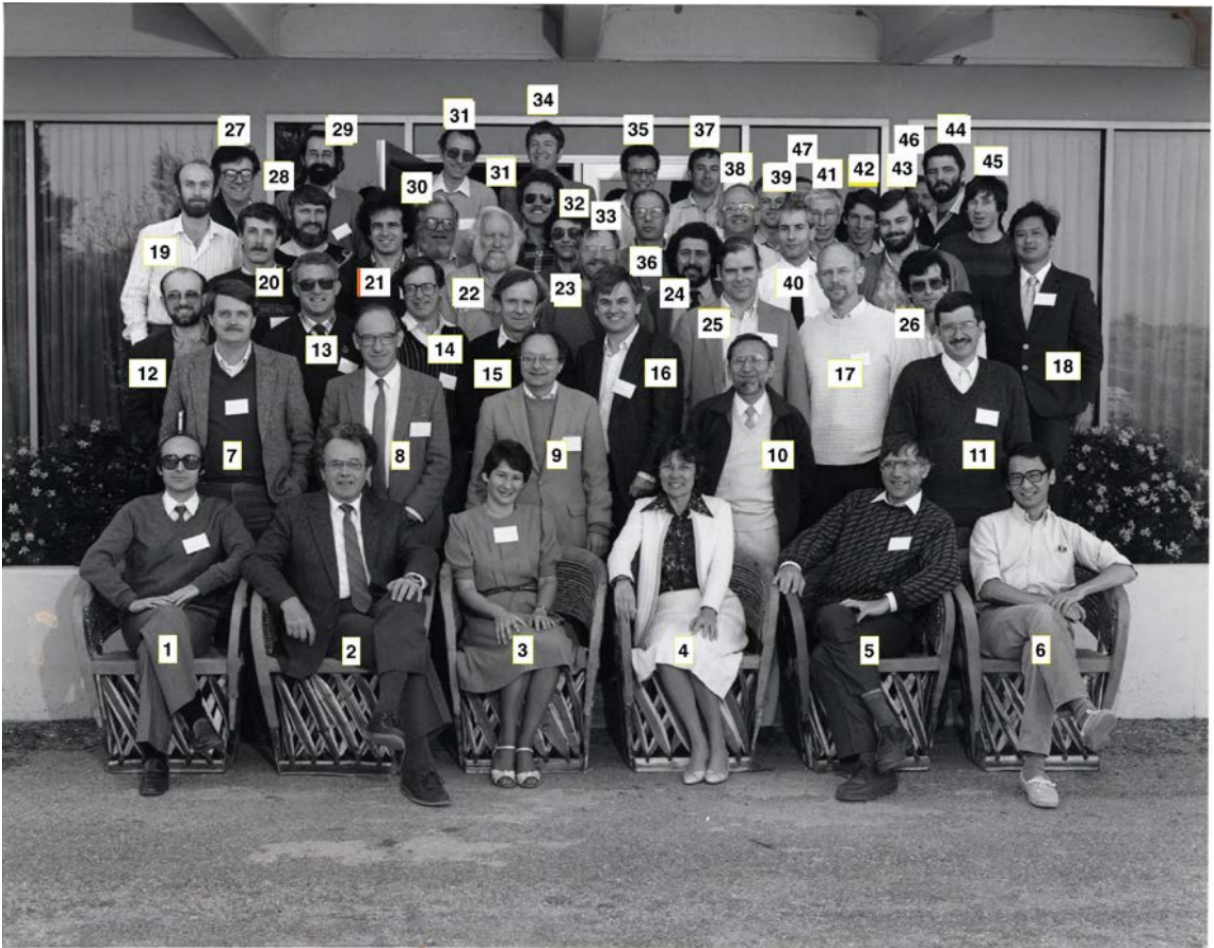

Figure 1: During the 1980s, the then very small optical interferometric community met regularly on both sides of the Atlantic, comparing strategies and early results. Photograph taken during the ESO-NOAO Joint meeting at Oracle (Arizona) in 1987. 1/ A. Lannes 2/ J. Beckers 3/ J Goad 4/ C. Roddier 5/ P. Léna 6/ M. Shao 7/H. McAllister 8/J. Davis 9/W. Bagnuolo 10/F. Roddier 11/C. Aime 12/ K.H. Hofman 13/D. McCarthy 14/D. Downes 15/ J. Baldwin 16/G. Weigelt 17/D. Currie 18/B. Qian 19/A. Greenaway 20/R. Howell 21/F. Merkle 22/R. Bates 23/M. Colavita 24/J. Christou 25/J. Fienup 26/C. Perrier 27/T. Cornell 28/C. Leinert 29/J.M. Mariotti 30/K. Hege 31/P. Nisenson 32/O. von der Lühe 33/E. Kibblewhite 34/S. Ridgway 35/A. Chelli 36/G. Aitken 37/E. Ribak 38/M. Dyck 39/M. Northcott 40/P. di Benedetto 41/J.C. Fontanella 42/G. Ayers 43/R. Petrov 44/J. Noordham 45/L. Koechlin 46/M. Coob 47/P. Connes. On this photo are missing some participants : R. Frieden, A. Labeyrie, R. Narayan, T. Readhead, W. Traub.

## 2. Recommended science cases

Optical interferometry has made substantial contributions to many scientific areas over the past decades. Not surprisingly, the greatest science impacts were achieved in domains that require high-angular resolution and high-precision measurements. While the scientific applications of such a unique technique are very wide, there is a consensus in the scientific community on a few science areas where the greatest science impacts from optical interferometers are expected to be realized today and in the near future. The most persistent trend is clearly the near-IR and mid-IR study of planet forming regions, which was also mentioned in the US Astro2010 Decadal survey report as the highest impact area for optical interferometry. The other main topics being clearly on a converging trend are stellar physics and the study of AGNs. In the following subsections, we give more details about each of these science topics.

## 2.1 Stars and stellar evolution

While the theory of stellar evolution was thought to be well established for single stars, asteroseismology recently revealed that the basic assumptions about interior rotation, and along with it chemical mixing and angular momentum transport, have major shortcomings (e.g., Chaplin & Miglio 2013 and Aerts 2015 for summaries for low-mass stars and Aerts et al. 2017 for intermediate-mass stars). A general conclusion is that unknown strong coupling between the stellar core and envelope must occur as stars evolve (Eggenberger et al 2017). The sample of stars with interior rotation from asteroseismology is currently limited in mass range, metallicities, and evolutionary stages. This will change soon thanks to the TESS and PLATO missions (Rauer et al. 2014, Ricker et al. 2016; launches in 2018 and 2026).

The quest to measure a stellar radius with high precision is more appealing than ever for all stars, but particularly those for which asteroseismology cannot be of help. This is the case for stars with considerable mass loss or accretion. The capacity to achieve a high-precision radius is particularly pertinent for stars in their formation process while they are forming their planetary systems. Even in the case of future successful asteroseismic applications of stars just before the hydrogen burning in their core (e.g. Zwintz et al. 2014 for a proof-of-concept), the availability of an interferometric radius during the various stages of stellar life would be a major asset to understand all phases of stellar evolution. Indeed, the combination of a high-precision angular diameter, distance (from Gaia), and spectroscopic effective temperature delivers a model-independent luminosity with far better predictive power for the evaluation and calibration of stellar evolution models than available presently. The addition of asteroseismic measurements of the interior properties, including rotation profiles, to an interferometric radius implies the capacity of high-precision inferences of key quantities of the interior physics of stars, decreasing drastically the model dependency of their aging (e.g., Huber et al. 2013). Hitherto the ages of stars in the formation process are not known, preventing the derivation of a relative timeline for the various disc properties that have been measured. The aging of stars in their formation process can be achieved from a combined asteroseismic and interferometric approach. It would allow defining a homogeneous star formation scenario from quasi model-independent relative aging.

Optical interferometers will also play a key role for the most massive stars in the Universe, which are continuously facing heavy mass loss since the start of their life. It was proven recently that more than 80% of all stars with masses above some 25 solar masses at birth occur in binaries or are the merger product of binary interaction (e.g., Sana et al. 2012). The first systematic high angular resolution survey to search for companions within a physical distance below some 200 AU seems to indicate that all O stars originate from binary formation channels. Hence stellar evolution theory for the most massive stars requires drastic improvements and interferometry has a major role to play here, as discussed by Sana et al. (2014). Dedicated combined interferometric and spectroscopic long-term monitoring is certainly the best and most efficient way to make progress in this area of stellar evolution, which is of prime importance for chemical galactic evolution. An extremely important step ahead could come from long-term time-resolved stellar imaging at visual wavelengths rather than in the infrared. Indeed, direct mapping of the surface of seismically and/or magnetically active regions on the surface of stars could reveal modest temperature or chemical spots as well as pulsational patches, revealing directly the spot and pulsation configurations (see e.g. Roettenbacher et al. 2016 and Roettenbacher et al. 2017). This would open up an entirely new research field: local seismology of (active) stars, following on local helioseismology (e.g., Gizon 2009). Interferometric observations at visual wavelengths is challenging but have been successfully achieved on the CHARA array (with the VEGA and PAVO combiners) and NPOI (CLASSIC and VISION combiners). For the VLTI, this is currently a technological challenge but the gain of achieving this in terms of scientific exploitation would be appreciable (see e.g., Stee et al. 2017).

## 2.2 Planet formation

Planet formation represents one of the major unsolved problems in modern astrophysics (Millan-Gabet et al. 2010). Planets are believed to form out of the material left over by the star formation process but the details on how it actually happens are still speculative and several theories exist. Constraining planet formation theories is a difficult task for a couple of reasons. On one hand, spectral energy distributions and spectroscopy alone do not uniquely constrain disc models. On the other hand, spatially resolving the inner disc region (interior to ~10 AU, most relevant in the context of planet formation) poses difficult observational challenges. This is where long-baseline interferometers have a key role to play. Indeed, direct imaging observations are crucial to break model degeneracies (e.g., Kraus et al. 2010) and can provide unexpected new results (see e.g. Creech-Eakman et al. 2010).

Recent advent in imaging capabilities at the VLTI and CHARA, sometimes combined with high-resolution spectroscopy, enabled direct and fundamentally new measurements of the inner region of protoplanetary discs: the radial location of the sources of continuum and line emission, gas chemistry and dust mineralogy, and surface brightness. However, major fundamental questions remain unanswered:

- The detailed structure and composition of the dust evaporation front, which is fundamental to the knowledge of the terrestrial planet formation zone.
- The density and temperature profiles of proto-planetary discs and how to explain the location/migration of gas giant planets or disc clearings.
- The connection between the disc and the star itself, particularly with regards to angular momentum transportation, magneto-hydrodynamics of accretion, and disc viscosity.

In the short-term, the second-generation instrument VLTI/MATISSE (Lopez et al. 2014) will improve our spatial coverage and imaging capabilities in the mid-infrared (L, M, and N spectral bands) by combining four telescopes instead of two previously with VLTI/MIDI. Single-baseline measurements often provide ambiguous interpretations, and MATISSE will play a crucial role to address pivotal questions about the inner regions of young circumstellar discs, such as those listed above. At CHARA, installation of adaptive optics and upgrades to the MIRCx combiner will improve the limiting magnitudes and give access to fainter young stellar objects. Next generation combiners like MYSTIC (K-band) and SPICA (visible, Mourard et al. 2017) are also being planned/discussed. NPOI will be adding 1-m telescopes to improve throughput. The Magdalena Ridge Observatory Interferometer (MROI) is under construction. In the long term, one of the most challenging goals is to probe planet-forming systems at the natural spatial scales over which material is being assembled (the so-called Hill Sphere, which delimits the region of influence of a gravitating body within its surrounding environment). The Planet Formation Imager project (PFI; see other paper in this volume) has crystallized around this challenging goal: to deliver resolved images of Hill-Sphere-sized structures within candidate planet-hosting discs in the nearest star-forming regions.

Finally, it is also worth mentioning the study of extrasolar planets as a persisting important goal of long-baseline interferometry. While young giant exoplanets are within reach of a high-contrast ground-based instrument (see e.g., description of the Hi-5 project in this volume), a space-based interferometer will be required to address the most fundamental questions such as the habitability of rocky exoplanets and the search for biosignatures (e.g., Cockell et al. 2009). There are however many technological challenges to overcome before launching such an ambitious instrument and a vigorous investment is needed (see other contribution in this volume).

## 2.3 Study of AGN

Starting from a few brightest objects about a decade ago, AGN have been observed with long-baseline infrared interferometers quite extensively over the last decade (Swain et al. 2003; Wittkowski et al. 2004; Jaffe et al. 2004; Meisenheimer et al. 2007; Tristram et al. 2007; Raban et al. 2009; Burtscher et al. 2009; Kishimoto et al. 2009; Pott et al. 2010; Kishimoto et al. 2011a,b; Weigelt et al. 2012; Hoenig et al. 2012, 2013; Burtscher et al. 2013; Tristram et al. 2014; Lopez-Gonzaga et al. 2014). We have now tens of AGN with interferometric size measurements in the infrared. So far, all these size measurements focus on the torus, the inner dusty region considered to surround mostly equatorially the central engine. One current finding is that the characteristic size over decreasing IR wavelengths from mid-IR to near-IR decreases quite fast, a little faster than the spatial resolution for a given baseline, at least in some objects. Since AGN observations are currently limited to baselines up to ~100 m, the same object is often more resolved in the mid-IR than in the near-IR. In fact, with the Keck Interferometer and VLTI (baseline 85-130m), objects have been relatively well resolved in the mid-IR for a significant number of cases, while only very partially resolved in the near-IR, except for one object. Based on the brightness of the objects in the AGN catalogue of Véron & Véron 2010, it is clear that the number of doable objects will explode with a small advancement of the current limiting magnitudes, which we definitely need. So far, these observations have mainly been limited to broad-band (continuum) observations, but high spectral resolution observations are being attempted in the near-IR as we discuss below.

In the mid-IR, a few objects have been relatively well resolved with the VLTI baseline lengths, and overall sizes and some radial structure have been measured for these and some more objects. However, the observations have been limited to those with two beams, meaning that we lack phases or more technically, we still have no closure phase information -- essentially, we do not have an "image" of each object yet. Differential phases over the mid-IR spectrum are being utilized, but the application is still limited and complicated. The next robust step forward is to get phase information with a ≥3 beam interferometer, leading to a first imaging. Based on the existing observations, we do have some morphological information, but this provides an even greater motivation to explore further. The two largest AGN in angular size on the sky, NGC1068 and Circinus, seem to show a two-component structure -- an equatorial one at several sublimation radii $R_{sub}$, and a polar elongation at a larger scale, around a few tens of $R_{sub}$ (Tristram et al. 2014; Lopez-Gonzaga et al. 2014). Somehow, this polar component is seen both in Type 2 and Type 1 objects (corresponding to edge-on and face-on objects, respectively; Hoenig et al. 2012, 2013). Naively, we would have expected to see a more equatorially elongated structure for the obscuring torus, but this is not the case, and interestingly the polar elongated component is even radiating more predominantly in the mid-IR. This could be an outflowing material, but at the same time, it seems to be participating in obscuring the central region, since the interferometric size measurements indicate an emissivity of a few tenths, which is expected from a directly illuminated, UV-optically-thick material. It is puzzling that this polar elongated structure is seen in a Type 1, face-on object, i.e. without obscuring the centre. Furthermore, in the two edge-on objects, NGC1068 and Circinus, the equatorial component shows quite a good correspondence with maser spots observed in the radio domain at least roughly in size and direction. It is well known that these spots rather suggest a warped structure, while UV/optical high-resolution images such as those with HST show a clear cone-like structure indicating a corresponding shadowing. With the mid-IR interferometric imaging, we will need to sharply show the structure that reconciles both requirements. That will potentially give us a clue on the nature of this obscuring material and the relation to the accretion flow.

In the near-IR, for Type 1 (supposedly face-on) AGN, which are currently the main objects we can observe interferometrically, it has been a bit more difficult to resolve the structure with 100 m baselines than in the mid-IR -- the wavelength dependence of the characteristic spatial scale often decreasing a little faster than $\propto \lambda$. The visibilities we observe in the near-IR with 100 m baselines for the brightest Type 1 AGN turn out to be $V^2 \sim 0.9$, thus we are only marginally resolving the structure. The situation with fainter ones would be even worse since they probably have a smaller angular size. What we definitely need is a longer baseline length, in order to unambiguously resolve the structure. At the moment, we believe that we are partially resolving a ring-like dust-emitting region, having a 'hole' due to the dust sublimation by the central engine's harsh heating. We will first be able to confirm this picture with 300m-class baselines with sufficient sensitivity. On the other hand, the central engine, the putative accretion disc, is believed to remain unresolved. However, the outermost part of the disc, where the mass accretion onto the disc truly occurs, is the region we poorly understand. As we approach these radii with the higher resolution, we can prove or disprove, and explore, how purely they remain to be unresolved for the first time. Before we reach this accretion region, we will actually go through the region mainly emitting broad emission lines. The kinematics and geometry of this region is very important as they are being used for estimating the mass of the central black hole. With a spectral resolution of ~100 km/s, detecting such kinematics in differential phase spectra will thus be very yielding. Such medium spectral resolution observations have already been shown to be feasible with the current instruments for one AGN, 3C273 (Petrov et al. 2013), but again, the baseline lengths seem to be still too short. However, a simple differential *visibility* spectrum over a broad emission line already turned out to be intriguing and puzzling -- the tentative result is that the overall size of the emission line region looks slightly larger than that of the dust-emitting region at least in this particular case. We should be able to follow up this issue over the next couple of years even with the current instrument, and the observations with longer baselines will surely be very decisive.

## 3. Summary

Optical long-baseline interferometry provides a unique and powerful resource for astrophysics. Current interferometric facilities have now reached a level of technical and operational readiness enabling scientific breakthroughs. The untapped potential of optical interferometry is however still immense. It is today the only technique capable to probe at optical wavelengths spatial scales at sub-milliarcsecond angular resolution, which will be beyond the ability of planned extremely large telescopes and indispensable to study planet formation, understand the fundamental physics of stars, and unravel the structure of AGN.

Acknowledgment: DD thanks the Belgian national funds for scientific research (FNRS). MK acknowledges support from JSPS under grant 16H05731.